\documentclass[journal=jctcce,manuscript=article]{achemso}
\usepackage{graphicx} % Required for inserting images
\usepackage{amsmath}
\usepackage{gensymb}
\usepackage{mhchem}
\usepackage{xcolor}
\usepackage{booktabs}
\usepackage{hyperref}
\usepackage{cleveref}

\colorlet{review}{black}
\colorlet{review2}{black}

\title{Analytic nuclear gradients including oriented external electric fields in a molecule-fixed frame}

\author{Duc Anh Lai}
\affiliation{Department of Chemistry, Southern Methodist University, Dallas, TX 75275, USA}
\author{Devin A. Matthews}
\affiliation{Department of Chemistry, Southern Methodist University, Dallas, TX 75275, USA}
\email{damatthews@smu.edu}

\begin{document}

\begin{tocentry}
\includegraphics[width=7.5cm]{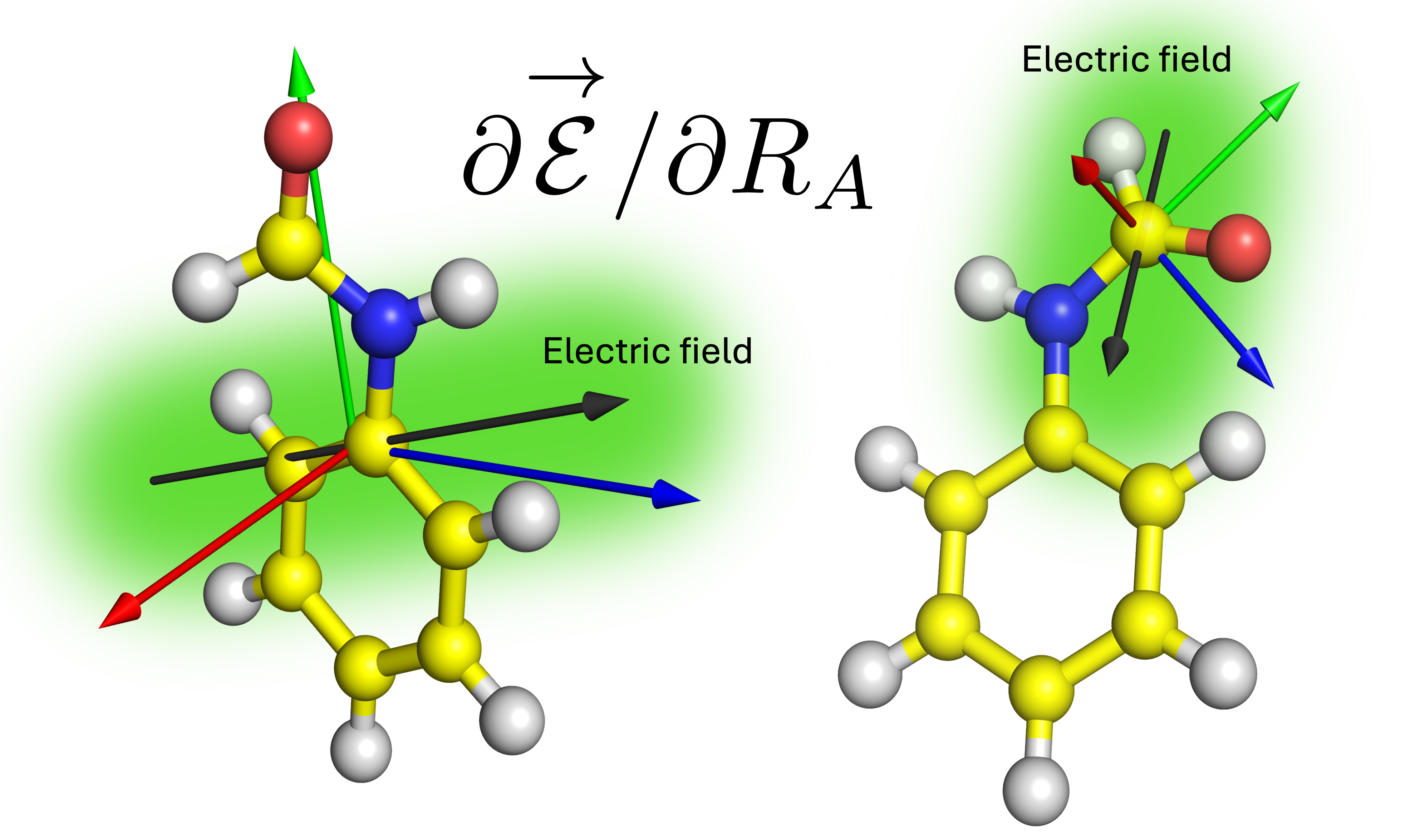}
\end{tocentry}

\begin{abstract}
Electric field–assisted chemistry has attracted much attention in recent years, particularly in the context of oriented external electric fields for controlling molecular structure and reactivity. Such fields have been explored in a wide range of applications, including switching materials, nanoparticles, controllable catalysts, medicines, and clinical therapies. 
% However, the use of fixed fields in the laboratory frame becomes ineffective for flexible molecules, as conformational changes can significantly alter their orientations. 
\textcolor{review}{However, the determination of fixed fields in the laboratory frame becomes ineffective for flexible molecules, as conformational changes can significantly alter the relative orientation between the applied field and molecular structure.} In this work, we propose two molecular reference frames—the principal axis frame and the local reference frame—to define oriented electric fields within the molecular framework. These coordinate systems powerfully eliminate ambiguities in the relative orientation between the applied field and the molecule. Analytic nuclear gradients in the presence of external electric fields are derived and implemented, with an initial application to field-dependent geometry optimizations of \emph{cis}- and \emph{trans}-formanilide. Analysis of the resulting field-induced equilibrium structures reveals distinct structural responses, validating the accuracy and robustness of the proposed formalism. The analytic gradient framework enables systematic investigations of molecular properties and reactivity under arbitrarily oriented electric fields, opening new opportunities for computational modeling and rational design in electric field-controlled chemistry. 
\end{abstract}

\section{Introduction}

The control of chemical structure and reactivity through external perturbations represents a central theme in modern chemistry. Examples include catalysts, pH, electromagnetic radiation, heat, as well as the application of magnetic and electric fields to manipulate intrinsic molecular properties. Numerous studies have utilized external fields to induce peculiar electronic properties in small molecules \cite{arabi_effects_2011, jankowska_electric_2015, alemani_electric_2006}, framework materials \cite{foroutan-nejad_dipolar_2016, chu_crystal-facet-controlled_2025, wang_asymmetric_2025, durholt_tuning_2019}, drug delivery systems \cite{mirvakili_wireless_2021, ge_drug_2012, darwish_multi-responsive_2014, byrne_local_2015}, proteins \cite{wang_effect_2014, jiang_effects_2019, toschi_effects_2009, ojeda-may_electric_2010}, and so on. In particular, external electric fields can redistribute molecular charge density, perturb potential energy surfaces, and modify electronic structure \cite{shaik_electric-field_2020}. The Stark effect, in which molecular energy levels shift under an external electric field, is a spectroscopic manifestation of the effect of an external field on molecules \cite{stark_observation_1913, stark_beobachtungen_1914, harmin_theory_1982}. Ultimately, external electric fields can modulate electrostatic chemical processes such as proton transfer \cite{lai_external_2010, arabi_effects_2011, kato_tunneling_2025, chen_enabling_2023, lai_external_2010, cassone_molecular_2021, geng_oriented_2018, jankowska_electric_2015}, chemical reactions \cite{wang_oriented_2019, wang_oriented-external_2018, bandrauk_effect_2004, ramanan_catalysis_2018, leonard_electric_2021, akamatsu_electric-field-assisted_2017, stuyver_electrophilic_2019, chen_enabling_2023, gorin_interfacial_2013, meir_oriented_2010, joy_oriented_2020, dutta_dubey_solvent_2020, shaik_structure_2018}, molecular assembly \cite{fujii_electric-field-controllable_2019, wang_effect_2014, jiang_effects_2019, toschi_effects_2009, ojeda-may_electric_2010, martin_electric_2018, kuang_molecular_2022, ray_effects_2024, arbeitman_sars-cov-2_2021}, molecular switching \cite{yu_controlling_2022, foroutan-nejad_dipolar_2016, clark_effect_2019, alemani_electric_2006, oklejas_electric-field_2008, fujii_electric-field-controllable_2019, han_electric-field-driven_2020, tang_electric-field-induced_2020, fruchtl_electronic_2023, tong_fast_2008, kempfer-robertson_protonation_2023},  and many others. 

Because an electric field is a vector quantity, molecular responses fundamentally depend on both its magnitude and orientation \cite{shaik_electric-field_2020, shaik_oriented_2016, shaik_structure_2018}. Previous studies have demonstrated pronounced directional effects on molecular structure and bonding \cite{clark_effect_2019, wang_effect_2014, jiang_effects_2019, ojeda-may_electric_2010, cassone_molecular_2021, pansini_molecules_2018, cartus_polymorphism_2023, shaik_structure_2018, arbeitman_sars-cov-2_2021, delley_vibrations_1998}. For example, Sowlati-Hashjin and Matta reported distinct variations in bond lengths, dipole moments, and vibrational frequencies of diatomic molecules under parallel and antiparallel fields \cite{sowlati-hashjin_chemical_2013}. Shaik's group demonstrated field-induced modulation of bond polarity and reactivity, including field-controlled catalysis and regioselectivity \cite{ramanan_catalysis_2018, shaik_electric-field_2020, stuyver_electrophilic_2019, meir_oriented_2010, shaik_oriented_2016, dutta_dubey_solvent_2020, shaik_structure_2018}. In addition, Wang et al. showed that oriented external electric fields (OEEFs) act as electric tweezers that intensively catalyze uncommon halogen substitution reactions \cite{wang_oriented_2019, wang_oriented-external_2018}. 

In fact, OEEFs arise naturally in many chemical and biological contexts. Materials such as ferroelectrics possess intrinsic charge separation that generates homogeneous electric fields at their surfaces, which can also be engineered or modulated by applied fields \cite{wang_electric-field-induced_2025}. Moreover, the oriented electric field associated with the resting potential of lipid bilayer membranes plays a crucial role in the transport of biomolecules across cellular membranes \cite{ermakov_electric_2023, zhang_interplay_2019}. Macromolecules containing ions and charged regions, such as proteins, lipids, and nucleic acids, exhibit intrinsic polarization that creates internal electric fields in their local environments \cite{suydam_electric_2006, sowlati-hashjin_electrostatic_2021, bim_local_2021, drobizhev_local_2021, siddiqui_designed_2023}. This principle underlies vibrational Stark spectroscopy and contributes to the regulation of biological mechanisms \cite{laberge_intrinsic_1998, fried_measuring_2015, eberhart_methods_2025, lehle_probing_2005, zheng_two-directional_2022}. In pharmaceutical science and biomedical engineering, external electric field pulses are exploited in various techniques such as electroporation \cite{gehl_electroporation_2003, kotnik_electroporation-based_2015, yarmush_electroporation-based_2014}, iontophoresis \cite{wang_wearable_2025, panchagnula_transdermal_2000}, electrical stimulation \cite{meng_electrical_2021, jing_study_2019}, and tissue engineering \cite{markx_use_2008, ryan_electric_2021, saliev_therapeutic_2014} to modulate drug delivery and cellular responses, thereby improving clinical efficacy and selectivity. 

Despite growing interest in field-assisted chemistry, most computational studies involving electric fields currently neglect molecular orientation effects. To the best of our knowledge, many computational chemistry programs employ electric fields in a laboratory-fixed frame (LF), in which the field orientation remains fixed with respect to the experimental setup rather than the molecular framework. While this approach is appropriate for rigid systems, it becomes unreliable when the field-relaxed structure leads to substantial reorientation. In such cases, the effective relative orientation between the molecule and the field becomes ill-defined.
\textcolor{review}{
As a consequence, in previous computational studies, this field-induced reorientation is either completely ignored or not treated appropriately. For example, many studies avoided the reorientation artifacts by aligning the field with the molecular dipole moment and implicitly assuming that the field does not alter the principal axis of rotation of the molecule \cite{rinconActivationSbondsElectric2016, shaik_structure_2018, ramanan_catalysis_2018, wang_oriented_2019, stuyver_electrophilic_2019}. Experimentally, however, the electric field can present as misaligned with respect to the principal axis or local electronic frame of the specific analyte of interest, such as in protein active sites \cite{zheng_enhanced_2023}, side-gate single molecular junctions \cite{yangUtilizationElectricFields2025}, supramolecular capsules, and zeolites \cite{welbornComputationalOptimizationElectric2018}. Alternatively, when the field is defined with respect to a specific internal coordinate (e.g., a reaction axis), geometry optimizations in internal coordinates can partially enforce a consistent field direction \cite{meir_oriented_2010, hanawayAutomatedVariableElectricField2023}. While effective for simple cases, such approaches may introduce artifacts or discontinuities in the field orientation.}
For the limitations above, a molecular frame (MF) description, in which the field follows the intrinsic molecular orientation, is therefore highly desirable for systematically studying OEEF effects on molecular systems. \textcolor{review}{Particularly, the use of OEEFs in the MF allows investigation of the response of a molecule to different field directions (that is, different external experimental conditions such as cage or active site geometry, interaction angle, crystal or grain orientation, etc.), thereby isolating intrinsic electric field effects from extrinsic rotational effects.}

An unambiguous coordinate frame consists of three mutually orthonormal unit vectors. In this work, we use $\{\mathbf{x,y,z}\}$ to denote three coordinate basis for the LF and $\{\mathbf{a,b,c}\}$ accordingly for the MF. One possible choice is to define an MF from intrinsic molecular properties, such as moments of inertia, quadrupole moment, g-tensor, or polarizability. These properties are represented by rank-2 tensors, which can be diagonalized to yield three mutually orthogonal eigenvectors that define a coordinate system. This choice of MF is of great significance when an effective description of molecular rotation is required as in vibrational spectroscopy, normal mode analysis, and thermodynamic partition functions. The eigenframe of the inertial tensor, commonly referred to as the principal axis frame (PAF), is the most widely exploited. The PAF is defined by three eigenvectors of the mass moment of inertia matrix, given by
\begin{align}
    \mathbf{I}_m&=\begin{bmatrix}
\sum_im_i(y_i^2+z_i^2)&-\sum_im_ix_iy_i&-\sum_im_ix_iz_i\\-\sum_im_iy_ix_i&\sum_im_i(x_i^2+z_i^2)&-\sum_im_iy_iz_i\\-\sum_im_iz_ix_i&-\sum_im_iz_iy_i&\sum_im_i(x_i^2+y_i^2)
\end{bmatrix}
\label{eq:Im}
\end{align}
where $m_i$ and $\{x_i,y_i,z_i\}$ are the mass and Cartesian coordinates of the $i$th atom, respectively. A significant advantage of the PAF is that it is a unique frame which conveniently describes rigid rotational motion of a molecule, and is therefore the default option in most theoretical quantum chemical calculations. 
\textcolor{review}{For the electric field simulations, the PAF can be used to model relaxation for functional groups in a complex system due to local electric fields, e.g., frustrated Lewis pairs \cite{ma_preferred_2024}, ion pairs \cite{luoSynergyLiberatedAnions2026}, substrates inside a cavity \cite{dubeyLocalElectricFields2022, welbornComputationalOptimizationElectric2018}, vibrational Stark effect probes \cite{fried_measuring_2015}, and others. Often in modeling such effects, it becomes either computationally efficient or even necessary to employ cluster models which isolate a particular substrate molecule or fragment. Providing a convenient, well-defined frame which maintains the physically relevant external field orientation while allowing for significant nuclear relaxation is facilitated through the use of the MF.}
There are advanced vibrational frames, such as the Eckart and Sayvetz frames, which extend the PAF by satisfying the Eckart conditions, and thus minimizing rotation-vibration coupling \cite{szalay_eckartsayvetz_2014}. These frames become standard in studies of high-level vibrational spectroscopy, rovibrational Hamiltonians, and kinetic energy operators in internal coordinates \cite{mellor_molecular_2022, lauvergnat_numerical_2016, sadri_rovibrational_2014, yurchenko__rotationvibration_2005}.

Alternatively, an MF can be constructed from specific atoms or bonds, which provides a so-called local reference frame (LRF). For example, in order to study the vibrational Stark effect spectroscopy of anisonitrile, one coordinate axis may be chosen along the cyanide (CN) bond, while a second axis is defined perpendicular to the benzene plane. The simplest setting of an LRF is from two non-parallel bonds: the $\mathbf a$ axis is projected along one bond, the $\mathbf b$ axis is aligned with the normal of the plane formed by the two bonds, and the $\mathbf c$ axis is conventionally defined by the cross product of the first two axes. The LRF is advantageous when a specific bond or functional group is of interest. Compared to the PAF, the LRF provides direct chemical interpretability and is therefore particularly useful for studying reaction dynamics and intermolecular interactions. \textcolor{review}{Furthermore, scanning tunneling microscope (STM) experiments and field-effect transistors (FET) can be modeled by using terminal atoms and a third reference atom in the LRF \cite{yangUtilizationElectricFields2025}.}
However, as implied by its name, the LRF lacks a global molecular perspective and is thus primarily suitable for the analysis of local properties. Other LRFs, e.g. defined by bond-angle bisector planes, ring planes, and other local geometric features are also possible.

By vertically stacking the three orthonormal axes of the MF, we obtain a unitary matrix $\mathbf U$ to transform an electric field vector between the MF and the LF, e.g.,
\begin{align}
    \varepsilon_{LF}&=\mathbf U\varepsilon_{MF}=
[\mathbf a\;\mathbf b\;\mathbf c]\varepsilon_{MF}
\end{align}
In the MF, once a reference vector $\mathbf p$ is chosen by a convention, the electric field vector can be determined by a rotation $\mathbf R$ of $\mathbf p$ as follows
\begin{align}
    \varepsilon_{MF}&=\mathbf R\mathbf p\Vert\varepsilon\Vert
\end{align}
where the double vertical notation $\Vert\cdot\Vert$ denotes the magnitude of the applied electric field. There are multiple ways to define a rotation matrix $\mathbf R$, including representations from quaternions, Euler angles, Davenport angles, and others. In practice, two elemental rotations are sufficient to achieve the target electric field vector. It is important to note that all rotations are carried out within the MF.

In this work, we derive and implement analytic nuclear gradients for correlated wavefunction methods in the presence of oriented external electric fields within both PAF and LRF. Using \emph{cis}- and \emph{trans}-formanilide as representative test systems, we demonstrate the reliability and numerical stability of the implementation through systematic geometry optimizations and potential energy surface analyses. \textcolor{review2}{In addition, we present a practical example of electric-field-induced isomerization to illustrate the importance of frame selection in electric field modeling. We show that a laboratory-fixed field can introduce artificial rotational effects that are inconsistent with experimentally constrained systems, whereas a molecular frame provides a physically meaningful description of the field response.}

\section{Theory}

\subsection{Analytic nuclear gradients}

The Hamiltonian in the presence of a homogeneous electric field is given by
\begin{align}
    \hat H&=\hat H_0-\hat\mu\cdot\varepsilon
\end{align}
where $\hat H_0$ is the field-free Hamiltonian, $\hat\mu$ denotes the total dipole moment operator, and $\boldsymbol{\varepsilon}$ is the applied vector electric field. The total energy of a molecule is obtained as the eigenvalue of the Schrödinger equation,
\begin{align}
    E&=\langle{\Psi|\hat H|\Psi}\rangle=E_0-\langle{\Psi|\hat\mu|\Psi}\rangle\cdot\varepsilon
\end{align}
where $\Psi$ represents the quantum-mechanical state of the field-perturbed molecule and $\mu=\langle\Psi|\hat \mu|\Psi\rangle$ is the total average molecule dipole moment.
%The nonlinear response of the energy to the applied field can be expressed as a power-series expansion,
%\begin{align}
%    E&=E_0-\mu\varepsilon-\frac12\alpha\varepsilon^2-\frac16\beta\varepsilon^3-\cdots
%\end{align}
%where $\mu$, $\alpha$, $\beta$, $\cdots$ correspond to the static dipole moment, polarizability, hyperpolarizability, and higher-order response properties, respectively. From this expansion, the leading contribution to the field-induced energy change originates from the dipole–field interaction, whereas higher-order terms rapidly diminish as the powers of field strength decrease exponentially.

The derivative of the energy with respect to a perturbation $\chi$ (e.g., nuclear displacement) is given by
\begin{align}
    E^\chi&=E_0^\chi-\langle{\Psi|\hat\mu|\Psi}\rangle^\chi\cdot\varepsilon-\langle{\Psi|\hat\mu|\Psi}\rangle\cdot\varepsilon^\chi
    \label{eq:grad}
\end{align}
where $X^\chi=\partial X/\partial\chi$.
The first term, $E_0^\chi$, represents the gradient of the field-free energy. The full formalism of $E_0^\chi$ depends on the underlying quantum-mechanical method and is beyond the scope of this work, although it does not depend directly on the perturbation of the external field or the dipole operator and includes the field only through trivial inclusion of the $-\hat{\mu}\cdot\varepsilon$ term in the Fock matrix. 
The influence of the perturbation directly on the coupling with the external electric field is contained in the last two terms of the \Cref{eq:grad}. 
\textcolor{review}{Because the field-dependent contributions are fully separated from the method-specific term $E_0^\chi$, the present formulation is inherently method-independent. In practice, these additional terms can be implemented as a modular extension on top of any existing field-free analytic gradient framework, provided that analytic gradients of the underlying theory are available.}

The second term describes the change in the molecular dipole moment with respect to nuclear displacements, while the third term describes the reorientation of the field within the reference frame as discussed below. The total dipole moment consists of nuclear and electronic contributions,
\begin{align}
    \hat\mu&=\hat\mu_n+\hat\mu_e = \sum_iZ_i\mathbf R_i -\mathbf{\hat r}
\end{align}
where $Z_i$ and $\mathbf R_i$ denote the charge and vector position of nucleus $i$, respectively, and $\hat{\mathbf r}$ is the electronic position operator. 

The derivative (block of the Jacobian) of the nuclear dipole moment with respect to $\mathbf R_i$ is,
\begin{align}
    \mu_n^{\mathbf R_i}&=Z_i I_{3\times 3}
\end{align}
Since the electronic position operator is independent of nuclear geometry, the derivative of the electronic dipole moment with respect to nuclear displacements involves only derivatives of the basis functions,
\begin{align}
    \mu_e^\chi&=-\sum_{\mu\nu}D_{\mu\nu}\bigg[\langle{\phi_\mu^\chi|\mathbf{\hat r}|\phi_\nu}\rangle+\langle{\phi_\mu|\mathbf{\hat r}|\phi_\nu^\chi}\rangle\bigg]
\end{align}
where $D_{\mu\nu}$ is the relaxed density matrix in the atomic-orbital basis $\phi$. Origin dependence of $\hat\mu_n$ and $\hat\mu_e$ is neglected, since the corresponding terms cancel in the total dipole moment for neutral systems.

The last term in the energy derivative is associated with the rotation of the molecular frame (MF) with respect to the laboratory frame (LF) induced by nuclear displacements. When the applied electric field is fixed in the LF, this term vanishes. Consequently, most current implementations of analytic field gradients in quantum-chemistry programs neglect this contribution.
The derivative of the electric field can be written as
\begin{align}
    \varepsilon^\chi&=\mathbf U^\chi \mathbf R\mathbf p\Vert\varepsilon\Vert
\end{align}
where $\mathbf U^\chi$ represents the derivative of the transformation matrix. In this work, we present expressions for $\boldsymbol{\varepsilon}^\chi$ in both the principal axis frame (PAF) and the local reference frame (LRF).

\subsection{Principal Axis Frame}

In the PAF, $\mathbf U^\chi$ is obtained from the derivatives of the eigenvectors of the moment-of-inertia matrix $\mathbf I_m$ (Eq. \ref{eq:Im}). Let $A$, $B$, and $C$ be the eigenvalues corresponding to the principal axes ${\mathbf a,\mathbf b,\mathbf c}$, respectively. Each eigenpair satisfies,
\begin{align}
    \big(\mathbf{I}_m-A\mathbf{I}\big)\mathbf{a}&=\mathbf0
\end{align}
Differentiation with respect to $\chi$ yields
\begin{align}
    \big(\mathbf{I}_m^\chi-A^\chi\mathbf{I}\big)\mathbf{a}+\big(\mathbf{I}_m-A\mathbf{I}\big)\mathbf{a}^\chi&=\mathbf0
\end{align}
Projecting onto $\mathbf b^T$ and using the orthonormality condition $\mathbf a^T\mathbf b=\delta_{ab}$ gives
\begin{align}
    \mathbf b^T\mathbf I_m^\chi\mathbf a+(B-A)\mathbf b^T\mathbf a^\chi&=\mathbf 0
\end{align}
Therefore,
\begin{align}
    \mathbf b^T\mathbf a^\chi&=\frac{\mathbf b^T\mathbf I_m^\chi\mathbf a}{A-B}
\end{align}
Since $\mathbf a^\chi$ is orthogonal to $\mathbf a$, it can be expressed as
\begin{align}
    \mathbf a^\chi&=\frac{\mathbf b\mathbf b^T\mathbf I_m^\chi\mathbf a}{A-B}+\frac{\mathbf c\mathbf c^T\mathbf I_m^\chi\mathbf a}{A-C}
    \label{eq:paf}
\end{align}
Derivatives of the other principal axes are obtained by cyclic permutation.

\Cref{eq:paf} becomes ill-defined when the moment of inertia is degenerate, which occurs in symmetric and spherical top molecules. However, external electric fields typically lower molecular symmetry, making degeneracies unlikely. If well-defined derivatives at high-symmetry points are required, then instead the relation $(U^T)^\chi U + U^TU^\chi=0$ may be used which suggests setting $U^\chi=0$ in the degenerate subspace to provide a minimal perturbation.

\subsection{Local Reference Frame}

In the LRF, the frame is defined using three noncollinear atoms with positions $\mathbf R_A$, $\mathbf R_B$, and $\mathbf R_C$. The first basis vector is
\begin{align}
    \mathbf c&=\frac{\mathbf u}{\Vert\mathbf u\Vert}
\end{align}
where $\mathbf u=\mathbf R_B-\mathbf R_A$. The second basis vector is chosen as the normal to the plane formed by the three atoms,
\begin{align}
    \mathbf b&=\frac{\mathbf u\times\mathbf v}{\Vert\mathbf u\times\mathbf v\Vert}
\end{align}
where $\mathbf v=\mathbf R_C-\mathbf R_A$. The third basis vector is
\begin{align}
    \mathbf a&=\mathbf b\times\mathbf c
\end{align}

Only derivatives with respect to the coordinates of atoms $A$, $B$, and $C$ contribute to $\mathbf U^\chi$. The derivative of $\mathbf c$ is obtained using the chain rule,
\begin{align}
    \mathbf c^\chi&=\frac{\partial\mathbf c}{\partial\mathbf u}\frac{\partial\mathbf u}{\partial\chi}=\frac{(1-\mathbf c\mathbf c^T)}{\Vert\mathbf u\Vert}\frac{\partial\mathbf u}{\partial\chi}
\end{align}
In particular,
\begin{align}
    \mathbf c^{\mathbf R_A}&=-\mathbf c^{\mathbf R_B} \\
    \mathbf c^{\mathbf R_B}&=\frac{(I-\mathbf c\mathbf c^T)}{\Vert\mathbf u\Vert}
\end{align}
and all other derivatives vanish. Similarly, the derivative of $\mathbf b$ is
\begin{align}
    \mathbf b^\chi&=\frac{\partial\mathbf b}{\partial(\mathbf u\times\mathbf v)}\frac{\partial(\mathbf u\times\mathbf v)}{\partial\chi} \nonumber \\
    &=\frac{(I-\mathbf b\mathbf b^T)}{\Vert\mathbf u\times\mathbf v\Vert}\frac{\partial(\mathbf u\times\mathbf v)}{\partial\chi} \\
    \mathbf b^{\mathbf R_A}&=- \mathbf b^{\mathbf R_B}-\mathbf b^{\mathbf R_C}\\
    \mathbf b^{\mathbf R_B}&=-\frac{(I-\mathbf b\mathbf b^T)}{\Vert\mathbf u\times\mathbf v\Vert}[\mathbf v]_\times \\
    \mathbf b^{\mathbf R_C}&=\frac{(I-\mathbf b\mathbf b^T)}{\Vert\mathbf u\times\mathbf v\Vert}[\mathbf u]_\times
\end{align}
which involves the skew-symmetric matrix
\begin{align}
    [\mathbf u]_\times &=\begin{bmatrix}
    0 &-u_z & u_y \\
    u_z& 0& -u_x \\
    -u_y & u_x & 0 
    \end{bmatrix}
\end{align}
and similarly for $\mathbf v$. Finally, the derivative of $\mathbf a$ is given by
\begin{align}
    \mathbf a^\chi&=\mathbf b^\chi \times \mathbf c+\mathbf b\times \mathbf c^\chi \\
    \mathbf a^{\mathbf R_A}&=- \mathbf a^{\mathbf R_B}-\mathbf a^{\mathbf R_C}\\
    \mathbf a^{\mathbf R_B}&=- \frac{\mathbf u^T\mathbf v }{\Vert\mathbf u\Vert \Vert\mathbf u\times\mathbf v\Vert} \mathbf b\mathbf b^T - \frac{\mathbf c \mathbf a^T}{\Vert\mathbf u\Vert} \\
    \mathbf a^{\mathbf R_C}&=\frac{\Vert\mathbf u\Vert}{ \Vert\mathbf u\times\mathbf v\Vert}   \mathbf b\mathbf b^T 
\end{align}
The LRF is free of degeneracies, in contrast to the PAF. However, it becomes undefined when the three reference atoms are collinear. Therefore, one needs to wisely select the three atoms used to define the LRF.

\section{Implementation}

Analytic gradients for OEEFs defined in the PAF and LRF were implemented to cooperate with the current PySCF codes.\cite{sun_pyscf_2018, sun_recent_2020} The implementation can be found on GitHub.\cite{git} Single-point SCF calculations involving an external electric field are initialized using an \texttt{RHFEField} instance. Electric field gradient scanners for geometry optimization were implemented in classes that mirror the standard PySCF gradient scanners. For example, the \texttt{EFieldCCSDGradients} class is used for CCSD geometry optimizations in the presence of an OEEF. By default, the script performs orientation of the field in the PAF. In this frame, the eigenvectors of the inertia tensor are sorted in descending order, such that the first principal axis corresponds to the largest eigenvalue. If three non-collinear atoms are specified, the LRF is used instead. Correctness and numerical accuracy of the implementation was confirmed by comparison with gradients obtained by finite differences of energies, with fields aligned manually to the changing PAF or LRF of the displaced points \textcolor{review}{(see data in the Supplementary Information). The differences between analytic and finite difference gradients are below $10^{-6}$~a.u. at the equilibrium geometry.}

Analytic nuclear gradients in the presence of OEEFS were applied to optimize the geometries of cis- and trans-formanilide. All quantum mechanical computations were carried out at frozen core CCSD/cc-pVTZ level\cite{purvis_full_1982,dunning1989a} using the PySCF package.

\section{Results}

\textcolor{review2}{\subsection{Geometry optimization of formanilide in molecular frames}}

\begin{figure}
    \centering
    \includegraphics[width=\linewidth]{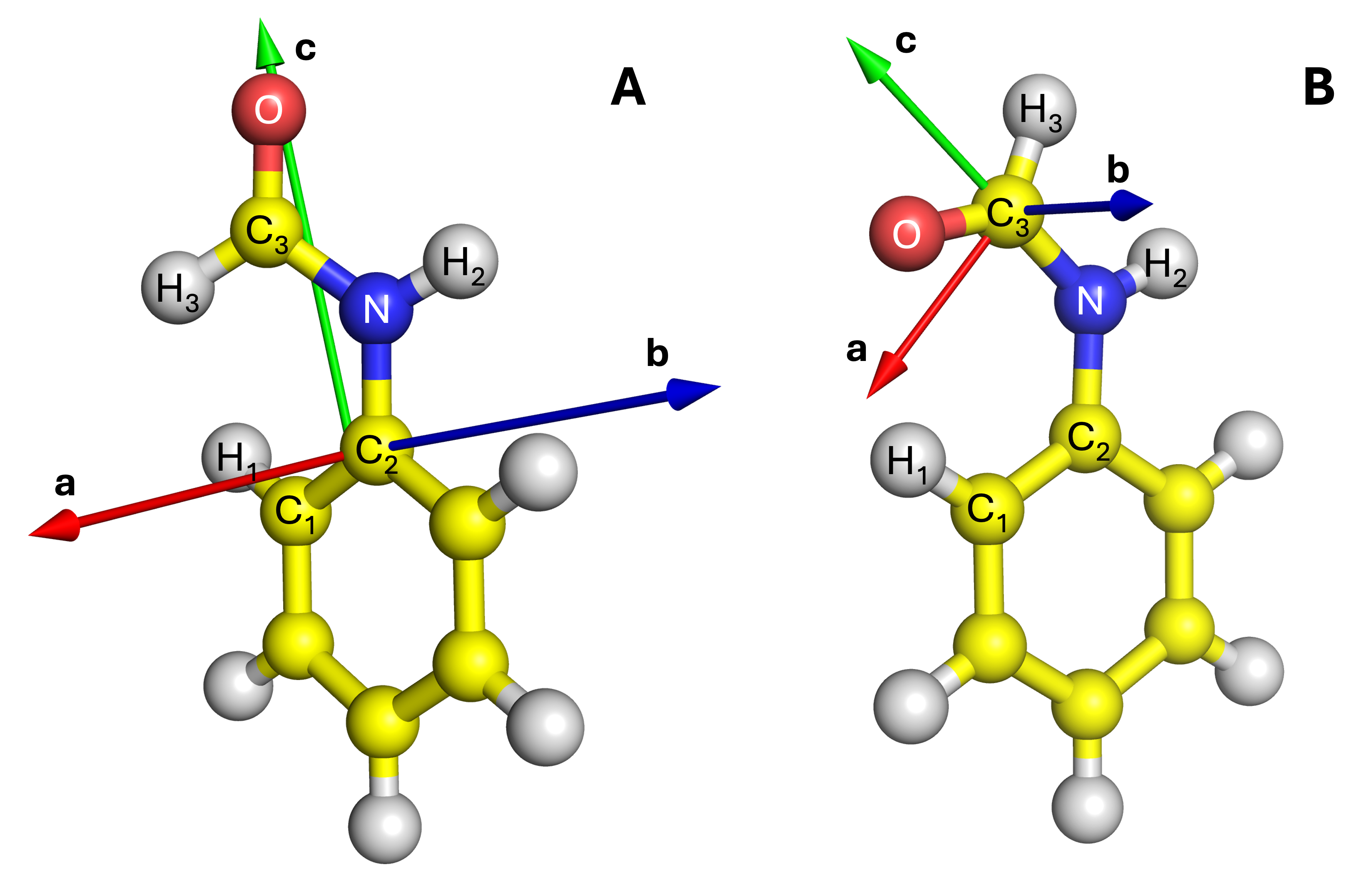}
    \caption{A) \emph{cis}-formanilide in the PAF. B) \emph{trans}-formanilide in the LRF.}
    \label{fig:axes}
\end{figure}

Formanilide is the simplest aromatic molecule containing a secondary amide group\break(\ce{R_1(CO)NHR_2}). The field-free molecule has been extensively investigated using both computational and experimental approaches. In particular, rotation about the amide \ce{C-N} bond gives rise to two distinct stable conformers, viz. the cis- and trans-isomers. The trans conformation adopts a syn-periplanar structure, while the cis conformation is twisted around the phenyl \ce{C-N} bond (\Cref{fig:axes}\textbf{B} and \Cref{fig:axes}\textbf{A}, respectively). Spectroscopic studies have confirmed that the trans isomer is more stable, resulting in a cis abundance of about 6.5\% \cite{blanco_conformational_2005, manea_conformations_1997, aviles_moreno_trans-isomer_2006, pasquini_trans-formanilide_2009}. However, cis/trans conformational isomerization of formanilide can be facilitated by hydrogen-bond networks formed by solvent molecules such as water \cite{pinacho_complete_2019}. In addition, the relative rotation between the amide and phenyl planes plays a critical role in stabilizing the cis conformer \cite{manea_conformations_1997}. Microsolvation studies have shown that the potential energy surface of amide torsional rotation is strongly perturbed by resonance-assisted hydrogen bonding \cite{pinacho_complete_2019, blanco_hydrogen-bond_2016, blanco_structure_2017, blanco_microsolvation_2006}. In light of these features, investigating the effects of an oriented external electric field on formanilide, particularly on its equilibrium dihedral torsions, is of considerable interest. Such studies can provide insight into intermolecular binding between monomers in peptide/protein structure and dynamics \cite{zhang_characterization_2011, pinacho_complete_2019, blanco_conformational_2005}.

Theoretical and experimental studies have revealed that the equilibrium conformations of formanilide result from competition of several intramolecular factors, including delocalization effects of the nitrogen lone pair into the aromatic ring and carbonyl group, as well as steric interactions between the amide hydrogens and phenyl hydrogens \cite{manea_conformations_1997}. In particular, resonance stabilization between the carbonyl bond and the nitrogen lone pair plays the dominant role and can extend into the $\pi$ system of the benzene ring. As a result, the trans-isomer adopts $C_s$ symmetry to maximize the overlap between phenyl-nitrogen and carbonyl-nitrogen conjugations, leading to enhanced stability. 

In contrast, steric repulsion between formyl and phenyl hydrogens in the cis form forces the phenyl ring out of plane, resulting in a strained non-planar conformation. In this case, amide delocalization is preserved, while the loss of nitrogen-phenyl conjugation is energetically compensated by steric relaxation. Furthermore, steric interactions between phenyl hydrogen and the oxygen atom in trans-formanilide are partially counteracted by a weak C–H···O hydrogen bond.

Because the cis-isomer adopts a non-planar conformation, the application of OEEFs induces distinct equilibrium \ce{C_1-C_2-N-C_3} dihedral angles, which largely affect the $\mathbf c$ principal axis in the PAF. Therefore, conformational changes of cis-formanilide under OEEFs were investigated using the PAF. In contrast, the effects of OEEFs on planar trans-formanilide were studied using the LRF.

\subsection{Geometry optimization of cis-formanilide in the PAF}

\Cref{tab:cis} summarizes the energies and geometrical parameters of cis-formanilide in the absence of a field and under external fields ranging from 0.01 to 0.05 a.u., applied along the $\mathbf c$ principal axis (\Cref{fig:axes}\textbf{A}). Two key energy differences were defined as follows: the $\Delta E_1$ difference corresponds to the energy barrier to rotation of the minimum-energy non-planar structure through a planar transition state, and the $\Delta E_2$ difference corresponds to the rotation barrier difference between two non-planar minima. In the latter case at zero external field, the two minima are equivalent and the transition state occurs precisely at $90\degree$; however in the presence of a field this symmetry is broken.

In the absence of an electric field, the coupled-cluster method accurately predicts the $\Delta E_1$ barrier (2.06 kJ/mol) compared to experimental value (1.82±0.02 kJ/mol \cite{blanco_conformational_2005}). The equilibrium \ce{C_1-C_2-N-C_3} dihedral angle is also well reproduced (36.3\degree\  versus 36.7\degree$\pm$2.7\degree \cite{marochkin_molecular_2013}). It is observed that ab initio methods, e.g., CC, MP2, HF, overestimate $\Delta E_1$ barriers and dihedral angles, whereas DFT methods underestimate them \cite{blanco_conformational_2005, marochkin_molecular_2013, manea_conformations_1997}. 

Under weaker electric fields of 0.01 and 0.02 a.u., the $\Delta E_1$ barrier increases to 4.19 and 5.15 kJ/mol, respectively. However, with further increases in field strength, the barrier decreases steadily and vanishes at 0.05 a.u. In contrast, the $\Delta E_2$ barrier initially decreases from 8.11 kJ/mol (field-free) to 1.29 kJ/mol at 0.03 a.u., before increasing sharply to 4.33 and 22.35 kJ/mol at 0.04 and 0.05 a.u., respectively. These variations indicate that the potential energy surface along the \ce{C_1-C_2-N-C_3} torsion is substantially flattened under moderate fields ($\Vert\varepsilon_{\textbf{c}}\Vert\leq0.04$ a.u.).

\begin{table}[]
\begin{tabular*}{\textwidth}{@{\extracolsep{\fill}} lcccccc @{}}
\toprule
                             & $\varepsilon_{\textbf{c}}=0$   & $\varepsilon_{\textbf{c}}=0.01$  & $\varepsilon_{\textbf{c}}=0.02$  & $\varepsilon_{\textbf{c}}=0.03$  & $\varepsilon_{\textbf{c}}=0.04$  & $\varepsilon_{\textbf{c}}=0.05$ \\ \midrule
$\Delta E_1$ (kJ/mol)   & 2.06  & 4.19    & 5.15    & 3.78    & 0.36    & 0.0     \\ 
$\Delta E_2$ (kJ/mol)   & 8.11  & 3.28    & 1.45    & 1.29    & 4.33    & 22.35   \\ 
\ce{C_1-C_2-N-C_3} ($E_2^\ddagger$) & 90.0  & 90.7    & 91.8    & 93.1    & 94.9    & 97.6    \\ 
\ce{C_1-C_2-N-C_3} (eq.)        & 36.3  & 44.9    & 50.1    & 50.0    & 24.9    & 0.0     \\ 
\ce{C_2-N}                         & 1.408 & 1.427   & 1.443   & 1.457   & 1.465   & 1.451   \\ 
\ce{C_3-N}                        & 1.367 & 1.348   & 1.332   & 1.320   & 1.312   & 1.322   \\ 
\ce{C_3-O}                        & 1.208 & 1.221   & 1.237   & 1.257   & 1.286   & 1.331   \\ 
\ce{H_1-H_3}                         & 2.262 & 2.436   & 2.571   & 2.605   & 2.334   & 2.394   \\ 
\ce{C_2-N-C_3}                      & 125.7 & 125.4   & 126.0   & 127.4   & 131.4   & 136.0   \\ 
\ce{C_1-C_2-N}                       & 121.3 & 120.6   & 120.5   & 120.6   & 122.7   & 125.1   \\ 
\ce{N-C_3-H_3}              & 112.9 & 112.4   & 112.0   & 111.8   & 112.1   & 112.7   \\ 
\ce{H_2-N-C_3-O}                  & 7.2   & 4.4     & 2.4     & 2.1     & 2.8     & 0.0     \\ \bottomrule
\end{tabular*}
\caption{Energies and structural parameters of cis-formanilide under OEEFs applied along the $\mathbf c$ axis in the PAF. Bond angles and dihedrals are in degrees and bond lengths are in \AA. Field strengths are in a.u.}
\label{tab:cis}
\end{table}

Notably, under applied fields, the ``perpendicular'' maximum involved in $\Delta E_2$ no longer occurs at exactly 90\degree\ as in the field-free case. Instead, the corresponding dihedral angle increases from 90.7\degree\ at 0.01 a.u. to 97.6\degree\ at 0.05 a.u. This asymmetric behavior can be attributed to field-amide interactions, as stabilization arises mainly from field-enhanced amide delocalization, which drives the \ce{C=O} dipole to align with the applied field. Consequently, the nitrogen atom is displaced from the phenyl plane, and the \ce{C_1-C_2-N-C_3} dihedral angle must increase to restore approximate equivalence on both sides of the phenyl ring. This behavior suggests a positive feedback mechanism for rotation about the \ce{C_2-N} bond in response to the external field.

\begin{figure}
    \centering
    \includegraphics[width=\linewidth]{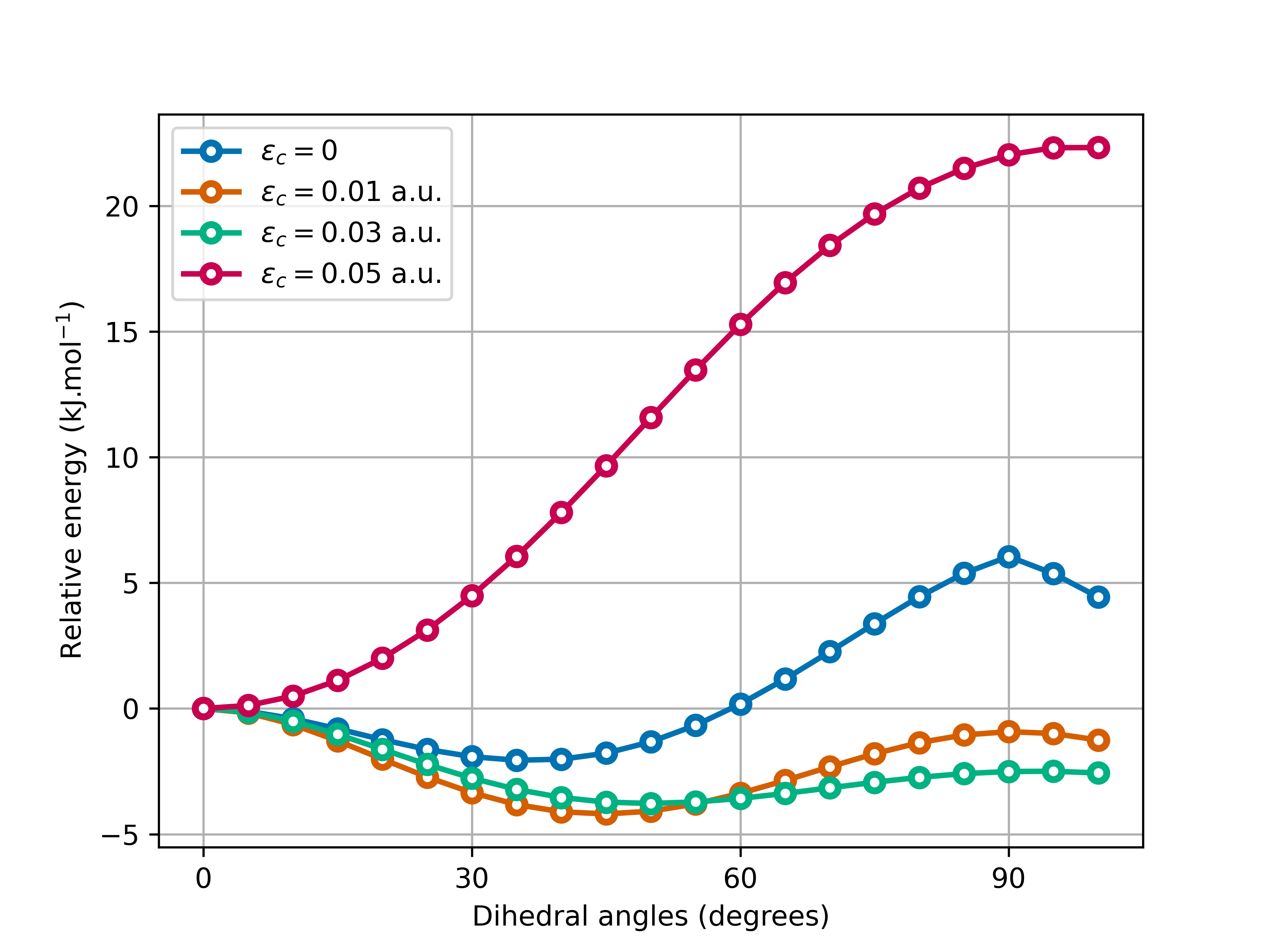}
    \caption{Potential energy surface along the \ce{C_1-C_2-N-C_3} dihedral angle of cis-formanilide under $\mathbf c$-aligned fields in the PAF.}
    \label{fig:pes}
\end{figure}

To further characterize the field-perturbed potential energy surface, PES scans along the \ce{C_2-N} rotation were performed at different field strengths (\Cref{fig:pes}). Similar to the $\Delta E_1$ barrier trends, the equilibrium dihedral angle increases from 36.3\degree\ (field-free) to 44.9\degree\ and 50.1\degree\ under weak fields, before decreasing to 24.9\degree\ at 0.04 a.u., and becoming planar at 0.05 a.u. These variations can be rationalized in terms of changes in electronic delocalization. Fields applied along the $\mathbf c$ axis hinder delocalization of the nitrogen lone pair into the aromatic system, as indicated by the lengthening of the \ce{C_2-N} bond, while enhancing amide charge separation, reflected in the shortening of the \ce{C_3-N} bond and lengthening of the \ce{C_3-O} bond. While weak fields (0.01--0.02 a.u.) disrupt phenyl-nitrogen conjugation, reducing the energetic benefit of planarity, steric repulsion between \ce{H_1-H_3} hydrogens becomes the dominant destabilizing factor, as evidenced by increases in the \ce{C_1-C_2-N-C_3} dihedral angle and the distance between the two hydrogens. In contrast, stronger fields induce an inverted phenyl-nitrogen resonance and favor co-planarity of the aromatic ring and amide group. To alleviate steric crowding between formyl and phenyl hydrogens, bond angles \ce{C_1-C_2-N}, 
\ce{C_2-N-C_3}, and \ce{N-C_3-H_3} expand. Additionally, stronger fields promote planarization of the amide dihedral angle (\ce{H_2-N-C_3-O}), further indicating enhanced amide conjugation.

Overall, the equilibrium structures of cis-formanilide in the presence of OEEFs defined in the global PAF are strongly driven by the field-amide dipole stabilization. These results confirm that the geometry optimization under OEEFs in the PAF yields physically meaningful structures and reliable energetic trends. 

\subsection{Geometry optimization of trans-formanilide in the LRF}

For trans-formanilide, an LRF was defined using the positions of the N, \ce{C_3}, and O atoms, with the $\mathbf c$ axis oriented along \ce{C_3-N} bond (\Cref{fig:axes}\textbf{B}). The planar configuration of field-free trans-formanilide allows us to separate the field effects into phenyl and amide contributions.

\begin{table}[]
\begin{tabular*}{\textwidth}{@{\extracolsep{\fill}} lllllllll @{}}
\toprule
 Field strength (a.u.) & $-0.04$       & $-0.03$     & $-0.02$    & $-0.01$    & $0.01$      & $0.02$      & $0.03$       & $0.04$       \\ \midrule
 \bf a-field \hfill \\ \midrule
\ce{C_1-C_2-N-C_3}   & 0.0        & 0.0       & 0.0      & 0.0     & 0.0      & 0.0     & 0.0       & 0.0      \\
\ce{C_2-N-C_3-O}   & 0.0        & 0.0       & 0.0      & 0.0     & 0.0      & 0.0     & 0.0       & 0.0      \\
\ce{H_2-N-C_3-O}  & 180.0   & 180.0  & 180.0 & 180.0 & 180.0 & 180.0  & 180.0   & 180.0   \\
\ce{C_2-N}        & 1.456    & 1.442  & 1.430 & 1.419 & 1.400  & 1.391   & 1.382    & 1.373    \\
\ce{C_3-N}        & 1.350   & 1.355 & 1.358 & 1.361 & 1.368  & 1.373   & 1.379    & 1.389    \\
\ce{C_3-O}        & 1.190    & 1.196  & 1.200 & 1.205 & 1.212  & 1.215   & 1.218    & 1.220 \\
\ce{N-C_3-O}     & 133.6    & 131.5  & 129.8 & 128.3 & 125.6  & 124.4   & 123.3    & 122.4   \\
\ce{H_3-C_3-O}     & 120.3    & 120.8  & 121.3 & 121.8 & 122.8  & 123.4   & 124.0   & 124.6  \\ 
field-phenyl & 0.0        & 0.0        & 0.0      & 0.0      & 0.0       & 0.0         & 0.0        & 0.0          \\ \midrule
\bf b-field \hfill \\ \midrule
\ce{C_1-C_2-N-C_3}   & -112.8       & -107.5     & -98.0     & -20.5     & 20.5       & 98.0        & 107.5        & 112.8        \\
\ce{C_2-N-C_3-O}   & 29.5         & 29.5       & 22.7      & 12.6      & -12.6      & -22.7       & -29.5        & 29.5         \\
\ce{H_2-N-C_3-O}  & 159.7    & 157.1  & 156.8 & 168.1 & -168.1 & -156.8  & -157.1   & -159.7   \\
\ce{C_2-N}        & 1.473    & 1.456  & 1.444 & 1.416 & 1.416  & 1.444   & 1.456    & 1.473    \\
\ce{C_3-N}        & 1.388    & 1.384  & 1.381 & 1.368 & 1.368  & 1.381   & 1.384    & 1.388    \\
\ce{C_3-O}        & 1.205    & 1.205  & 1.205 & 1.208 & 1.208  & 1.205   & 1.205    & 1.205    \\
\ce{N-C_3-O}     & 126.4    & 126.1  & 125.8 & 126.9 & 126.9  & 125.8   & 126.1    & 126.4    \\
\ce{H_3-C_3-O}     & 121.6    & 122.1  & 122.5 & 122.3 & 122.3  & 122.5   & 122.1    & 121.6  \\
field-phenyl & 1.8          & 1.7        & 3.1       & 73.5      & 73.5       & 3.1         & 1.7          & 1.8          \\
\midrule
\bf c-field \hfill\\ \midrule
\ce{C_1-C_2-N-C_3}   & 0.0        & 0.0       & 0.0      & 0.0     & 0.0      & 0.0     & 0.0       & 0.0      \\
\ce{C_2-N-C_3-O}   & 0.0        & 0.0       & 0.0      & 0.0     & 0.0      & 0.0     & 0.0       & 0.0      \\
\ce{H_2-N-C_3-O}  & 180.0   & 180.0  & 180.0 & 180.0 & 180.0 & 180.0  & 180.0   & 180.0   \\
\ce{C_2-N}        & 1.336    & 1.362  & 1.381& 1.396 & 1.421  & 1.432   & 1.441    & 1.450   \\
\ce{C_3-N}        & 1.536    & 1.447  & 1.410 & 1.385 & 1.348  & 1.335   & 1.325    & 1.321    \\
\ce{C_3-O}        & 1.171   & 1.186  & 1.193 & 1.201 & 1.217  & 1.229   & 1.243    & 1.263    \\
\ce{N-C_3-O}     & 118.9    & 121.1  & 123.2 & 125.1 & 128.6  & 130.4   & 132.4    & 134.6    \\
\ce{H_3-C_3-O}     & 130.0    & 126.4  & 124.7 & 123.4 & 121.3  & 120.2  & 119.0    & 117.5\\
field-phenyl & 0.0        & 0.0        & 0.0      & 0.0      & 0.0       & 0.0         & 0.0        & 0.0          \\ \bottomrule

\end{tabular*}
\caption{Structural parameters of \emph{trans}-formanilide under OEEFs applied along the three axes of the LRF.}
\label{tab:trans}
\end{table}

\Cref{tab:trans} summarizes the structural parameters of trans-formanilide under $\mathbf a$, $\mathbf b$, and $\mathbf c$ fields ranging from -0.04 to 0.04 a.u. Application of $\mathbf a$ and $\mathbf c$ fields, which lie within the molecular plane, preserves planarity, as reflected in the \ce{C_1-C_2-N-C_3}, \ce{C_2-N-C_3-O}, and \ce{H_2-N-C_3-O} dihedral angles. These fields primarily interact with the amide dipole and stabilize amide delocalization. 

Positive $\mathbf c$ fields enhance amide conjugation, shortening the \ce{C_3-N} bond from 1.368 \AA\ to 1.321 \AA\ and lengthening the \ce{C_3-O} bond from 1.212 \AA\ to 1.263 \AA\ as the field increases from 0.01 to 0.04 a.u. Conversely, negative $\mathbf c$ fields weaken amide conjugation, leading to elongation of the \ce{C_3-N} bond (1.385--1.536 \AA) and shortening of the \ce{C_3-O} bond (1.201--1.171 \AA). Notably, the \ce{C_3-N} bond approaches the dissociation limit when $\varepsilon_\textbf{c}\sim-0.04$ a.u. $\mathbf c$ fields also promote alignment of the amide dipole, mainly arising from the \ce{C=O} bond, as reflected in changes in the \ce{N-C_3-O} and \ce{H_3-C_3-O} angles.

\begin{figure}
    \centering
    \includegraphics[width=\linewidth]{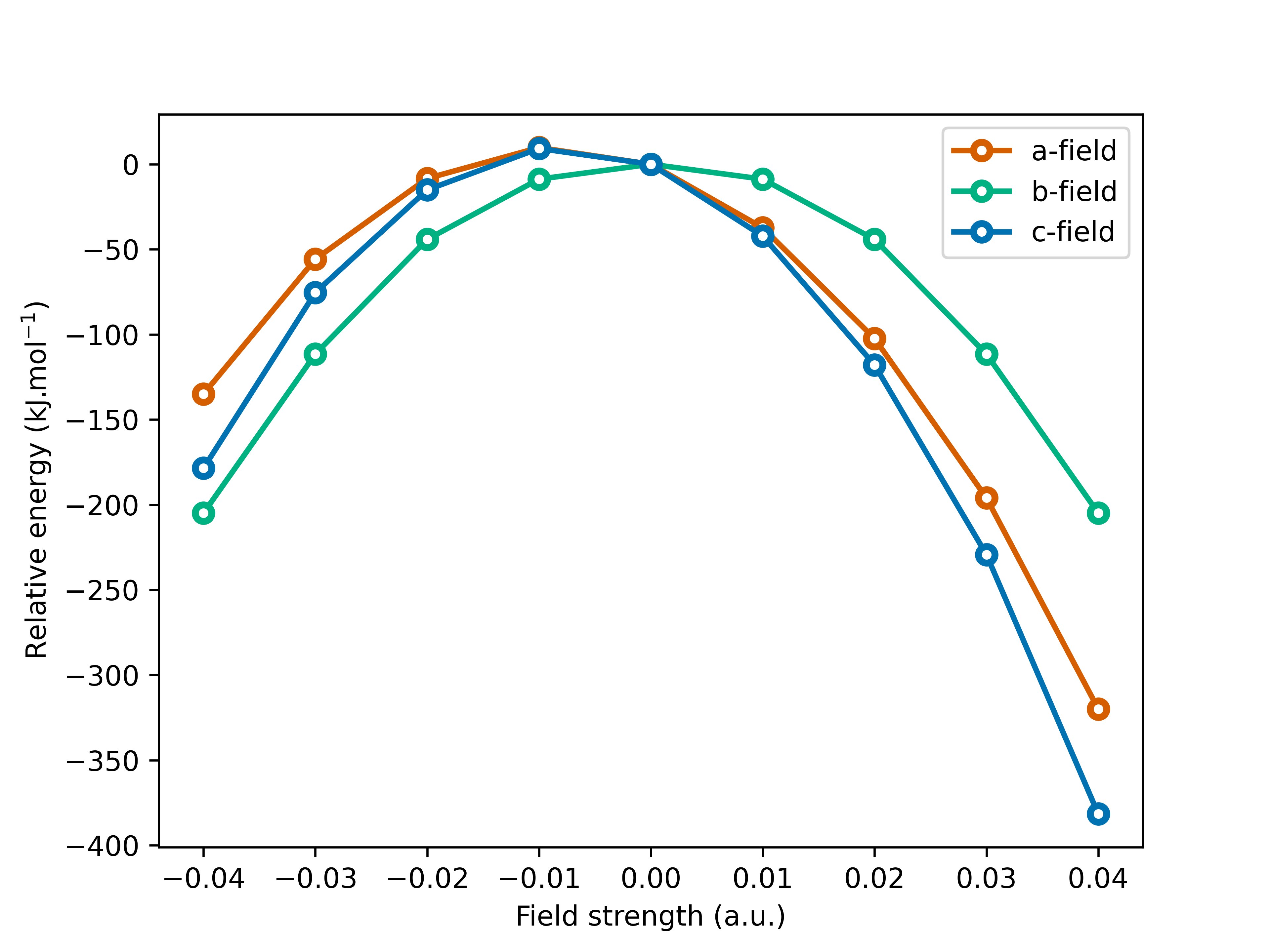}
    \caption{Total energies of trans-formanilide under OEEFs in the LRF.}
    \label{fig:energy_trans}
\end{figure}

The induced stabilization of $\mathbf a$-aligned fields mainly originates from field-dipole interactions, as reflected in reorientation of the \ce{C=O} bond. Specifically, the \ce{N-C_3-O} angle decreases from 133.6\degree\ to 122.4\degree\, while the \ce{H_3-C_3-O} angle increases from 120.3\degree\ to 124.6\degree\, improving alignment with the applied field. In comparison to $\mathbf c$ fields, $\mathbf a$ fields induce much weaker delocalization effects. Both \ce{C_3-N} and \ce{C_3-O} bonds increase slightly with $\mathbf a$-field strength, from 1.350 to 1.389 \AA\ and from 1.190 to 1.220 \AA, respectively. Consequently, energy relaxation under $\mathbf c$ field is consistently larger than under $\mathbf a$ fields (\Cref{fig:energy_trans}). In addition, the \ce{C_2-N} bond length increases under $\mathbf c$ fields but decreases under $\mathbf a$ fields of comparable magnitude.

\begin{figure}
    \centering
    \includegraphics[width=\linewidth]{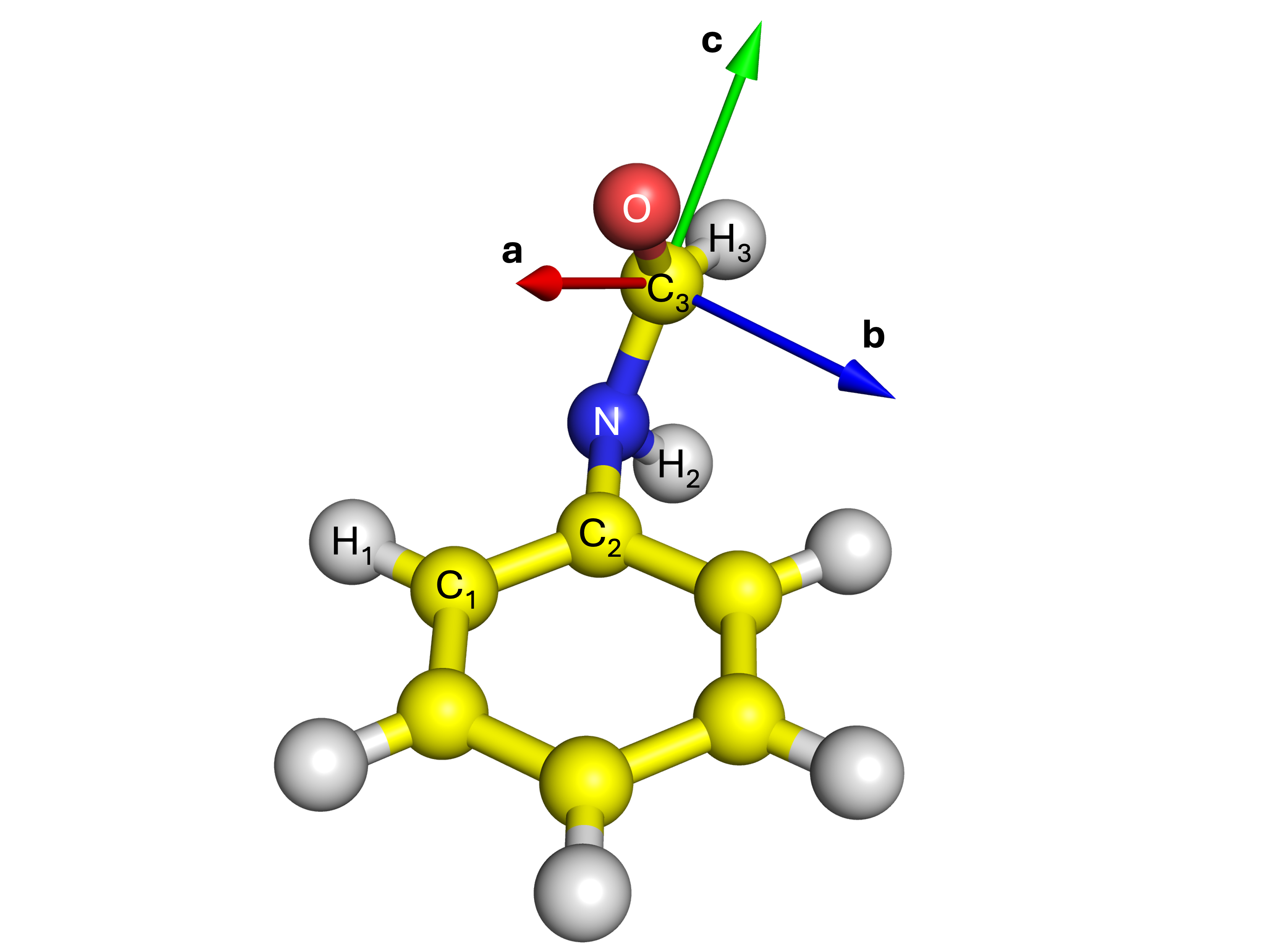}
    \caption{Out-of-plane twisting of trans-formanilide phenyl ring due applied $\mathbf b$-aligned field of 0.04 a.u.}
    \label{fig:twisted-phenyl}
\end{figure}

In contrast, $\mathbf b$ fields cause twisting between the phenyl ring and amide group, breaking $C_s$ symmetry (\Cref{fig:twisted-phenyl}). Both energies and geometries are symmetric about zero field due to the planar equilibrium structure. Because $\mathbf b$ fields are fixed perpendicular to the amide plane, they have minimal influence on amide stabilization, as indicated by nearly constant \ce{C_2-N}, \ce{C_3-N}, and \ce{C_3-O} bond lengths and associated bond angles. Instead, $\mathbf b$ fields mainly interact with the phenyl ring, driving reorientation of the aromatic system. This consequence reflects the nature of the LRF in which the relative angle between the applied field and amide plane remains unchanged. While the \ce{C_2-N} bond remains in the phenyl plane, twisting is reflected in increased \ce{C_1-C_2-N-C_3} and \ce{C_2-N-C_3-O} dihedral angles as $\varepsilon_\textbf{c}$ increases from 0.01 to 0.04 a.u. Consequently, the phenyl ring aligns progressively with the applied field, reducing the field-phenyl angle from 73.5\degree\ (0.01 a.u.) to 1.8\degree\ (0.04 a.u.). This behavior arises from the substantial in-plane polarizability of benzene, which enables strong induced dipole formation under $\mathbf b$ fields, as evidenced from the negligible change in the angle between the applied field and the phenyl plane at field strength 0.03 a.u. and 0.04 a.u.
 
Moreover, $\mathbf b$ fields distort the amide group from planarity, producing a gauche-like conformation, as reflected in deviations of the \ce{H_2-N-C_3-O} dihedral angle. This distortion results from opposing responses of the NH and CO fragments to the applied field, consistent with charge separation within the amide moiety. These results further validate the robustness of the analytic gradients within the LRF framework. 

\textcolor{review2}{\subsection{Field-induced keto-enol tautomerization}}

\textcolor{review2}{
While the PAF provides a conceptual framework for analyzing electric field effects relative to the intrinsic molecular geometry, the LRF more closely reflects many experimental configurations. In numerous electric field-driven experiments, including single-molecule junctions \cite{starrGoldCarbonContacts2020, diefAdvancesSinglemoleculeJunctions2023, casaliniSelfassembledMonolayersOrganic2017, metzgerUnimolecularElectronics2015, xiangMolecularScaleElectronicsConcept2016}, scanning tunneling microscope (STM) break-junctions \cite{xuMeasurementSingleMoleculeResistance2003, alemani_electric_2006, tang_voltage-driven_2023}, and nanoscale quasi-plate capacitors \cite{gorin_interfacial_2013, gorinElectricFieldInduced2012}, part of the molecule is mechanically or chemically constrained, such that its orientation remains fixed with respect to the applied field. In these experiments, the electric field is considered to maintain a constant orientation relative to a set of anchored atoms. Consequently, geometry optimizations performed in a fixed laboratory frame, in which positions of all atoms are freely optimized, may not accurately represent the experimental conditions. 
}

\textcolor{review2}{
To illustrate this limitation, we examine the field-induced keto-enol tautomerization of butan-2-one $\rightleftharpoons$ (\emph{Z})-but-2-en-2-ol. Electric-field control of keto-enol equilibrium has been extensively studied using STM techniques, where the field can be employed to modulate tautomeric stability \cite{tang_voltage-driven_2023, adijiang_regulating_2025}. To mimic such experiments, the terminal carbon atoms (C1 and C4) are anchored at the metal electrodes. The LRF is therefore defined using the two terminal carbons and the oxygen atom, and the external electric field is constrained to remain parallel to the C1--C4 axis throughout the optimization. In contrast, conventional LF calculations allow unrestricted molecular rotation, with the field referenced only along the initial molecular orientation. 
}

\begin{figure}
    \centering
    \includegraphics[width=\linewidth]{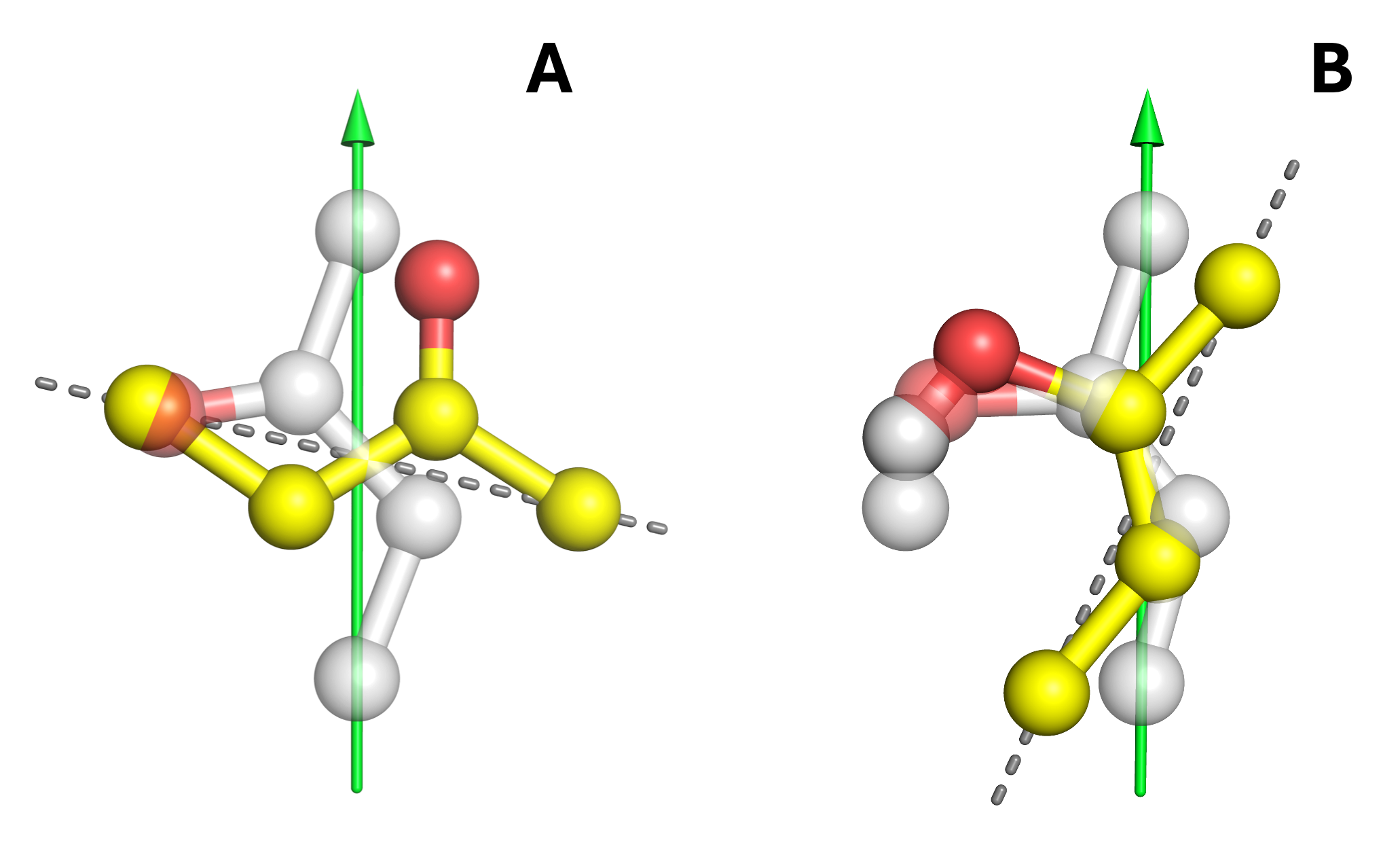}
    \caption{\textcolor{review2}{Geometry optimizations for the keto tautomer (A) and the enol tautomer (B) performed in the LF (yellow) exhibit artificial field-induced molecular rotation relative to optimizations performed in the LRF (white). The electric field vector (green solid line) forms angles of $102^\circ$ and $25^\circ$ with the C1--C4 axis (gray dashed line) for the keto and enol tautomers in the LF optimizations, respectively.}}
    \label{fig:enol-keto-mol}
\end{figure}

\textcolor{review2}{
Optimized keto and enol structures under an electric field strength of 0.01 a.u. are shown in \Cref{fig:enol-keto-mol}. For the keto tautomer, LF optimization leads to substantial molecular reorientation, resulting in perfect alignment of the applied field with the \ce{C=O} bond which is associated with the dipole moment of the conformation. Such rotation is energetically favorable in the LF because the entire molecule is free to reorient to maximize the alignment between molecular dipole moment and the applied field. In contrast, the LRF maintains the experimentally relevant condition in which the field remains fixed relative to the anchored termini. As a result, the optimized keto structure differs markedly from the LF geometry, in particular, the C1--C4 axis forms an angle of approximately $102^{\circ}$ with the field direction in the LF geometry (\Cref{fig:enol-keto-mol}\textbf{A}, this is reported at $>90^\circ$ considering directionality of the LRF). The artificial rotation is less pronounced for the enol isomer because the largest component of the permanent dipole moment is already approximately aligned with the C1--C4 axis, reducing the torque for field-induced rotation. Consequently, the angle between the C1--C4 axis and the applied field in the LF-optimized geometry is only about $25^{\circ}$ (\Cref{fig:enol-keto-mol}\textbf{B}). Overall, LF optimizations for both tautomers introduce a degree of artificial molecular reorientation that is absent in the experimental setup represented by the LRF. 
}

\begin{figure}
    \centering
    \includegraphics[width=\linewidth]{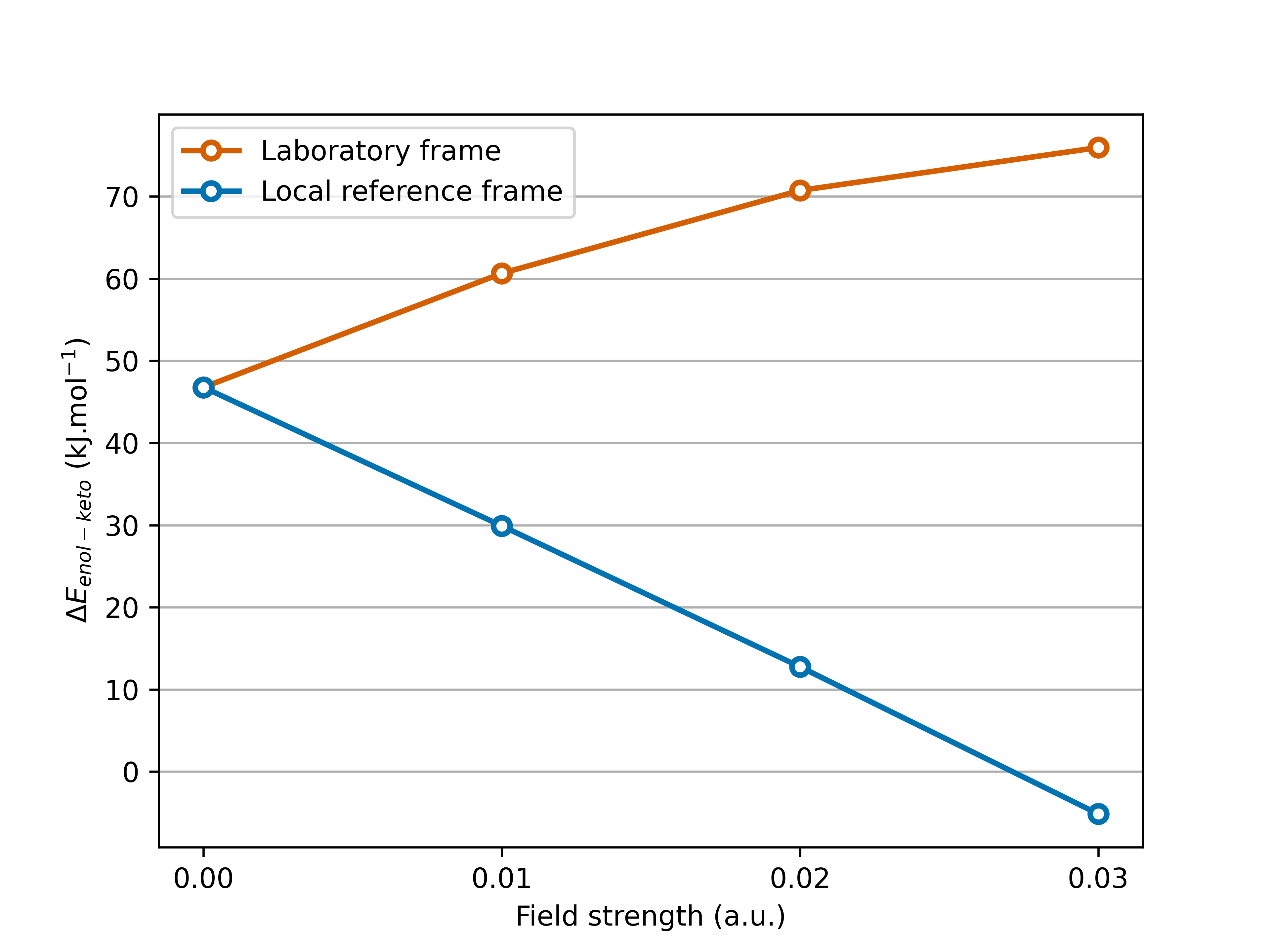}
    \caption{\textcolor{review2}{Relative energy between enol and keto tautomers ($\Delta E_{enol-keto}=E_{enol}-E_{keto}$) under field strengths of up to 0.03 a.u. in the LF and the LRF.}}
    \label{fig:enol-keto-energy}
\end{figure}

\textcolor{review2}{
Moreover, the choice of coordinate systems also has a significant impact on the relative energy between tautomers (\Cref{fig:enol-keto-energy}). In the absence of an electric field, the keto form is more stable than the enol tautomer by 46.8 kJ/mol, which is consistent with experimental observations that the enol form rapidly and completely converts to the more stable ketone \cite{turecek_unstable_1988, grajales-gonzalez_theoretical_2018}. Application of an external electric field alters this equilibrium through distinct responses of the dipole moment. In the LF, unrestricted rotation enables each tautomer to orient its dipole favorably with respect to the applied field. This rotational effect induces additional stabilization that would not be available in a constrained manner. Because the keto tautomer possesses a larger dipole component along the field direction after reorientation, it is stabilized greater than the enol tautomer. Consequently, the energy difference between the two conformations increases nonlinearly as the field strength increases up to 0.03 a.u. On the other hand, the LRF prevents artificial alignment and therefore isolates the intrinsic response of the molecular structure to the electric field under experimentally relevant conditions. Within the same field strength range, the LRF enol-keto energy gap decreases approximately linearly, consequently shifting the equilibrium toward the enol tautomer. Notably, at the field strength of 0.03 a.u., the relative stability is reversed ($\Delta E_{enol-keto}=-5.1$ kJ/mol), indicating the enol tautomer is more stable. 
The electric-field-induced reversal of enol-keto equilibrium was experimentally confirmed in two-terminal junction systems \cite{tang_voltage-driven_2023, adijiang_regulating_2025}.
These results demonstrate that modeling electric fields in the laboratory frame incorrectly describes the response to an external oriented electric field, leading to conclusions that are inconsistent with the physical constrains of realistic experiments.
}

\section{Conclusions}

Oriented external electric fields are prevalent in chemistry. Yet, understanding of oriented field effects on the geometries of chemical systems still remains largely unknown. The lack of analytic gradients in optimization frameworks within molecular frames has limited the comprehensive scale of electric field computational studies. In this work, we presented the analytic nuclear gradients for ab-initio wavefunction methods in the presence of oriented electric fields within both principal axis and local reference frames. This development enables correct geometry optimization and potential energy surface exploration under external fields in molecule-fixed coordinates. 
\textcolor{review}{Notably, this framework is particularly advantageous for model systems in which the field orientation is imposed as an intrinsic part of the molecular description, allowing oriented external fields to be treated consistently within the model itself. Moreover, the implementation of electric-field contributions to the gradients is method-compatible and can be readily interfaced with any level of electronic structure theory for which field-free nuclear gradients are available, including \textit{ab initio} wavefunction and density functional methods.}

 As an illustration, geometry optimizations of cis- and trans-formanilide were performed under the application of different field directions and strengths. Analyses from optimized structures reflected the nature of the computational coordinate frame. In particular, applying the field along the $\mathbf c$-axis in the PAF of cis-formanilide results in distinct torsional minima and energy barriers, as the structure was mainly driven by the interaction between the applied field and the $\pi$-conjugation system across the entire molecule. Whereas, computations in the LRF of trans-formanilide revealed the individual field-induced stabilization of either the phenyl ring (\textbf{b}-axis) or amide structure (\textbf{a}- and \textbf{c}-axes). Especially, applying in-plane fields maintained $C_s$ symmetry and mainly caused the distortion in amide delocalization. In contrast, perpendicular fields enforced an alignment of the phenyl to the applied field due to the in-plane polarizability of the benzene. This result may reveal electric field-induced structural changes in peptides and proteins, in which the peptidic linkage is relatively stiff but the side chain (e.g. phenyl ring) is more flexible.

\textcolor{review2}{In addition, we demonstrated that the choice of reference frame can qualitatively affect the predicted response of molecules to external electric fields. Using a keto-enol tautomerization in a molecular junction as an illustrative example, we showed that laboratory-frame optimizations allow artificial field-induced molecular rotation, leading to geometries and energetics that may not reflect experimentally constrained conditions. In contrast, the local reference frame preserves a fixed field orientation relative to the molecular framework and therefore provides a physically meaningful description of electric-field effects in STM and other single-molecule junction experiments. This result highlights the importance of molecular-frame nuclear gradients for accurately modeling electric-field-driven processes in constrained molecular systems.}
 
Together, the field-dependent optimizations yield smooth, physically consistent structural responses across a wide range of field strengths and orientations. The systematic and stable behavior observed in all cases 
\textcolor{review}{as well as the negligible deviation between analytic gradients and finite differences} 
confirms the correctness and numerical robustness of the present implementation. Integration of the analytic field-dependent gradients into PySCF provides a general platform for future investigations of electric field effects on molecular structure, spectroscopy, and properties. 

\begin{acknowledgement}
This work was supported in part by the US National Science Foundation (grant CHE-2143725) and by the US Department of Energy (grant DE-SC0022893). Computational
resources for this research were provided by SMU’s O’Donnell Data Science and Research
Computing Institute.
\end{acknowledgement}

\begin{suppinfo}
A Supplemental Information file (.xlsx) is available and contains all computed energies, optimized geometries, \textcolor{review}{and element-wise comparisons between finite differences and analytic gradients for optimized geometries}.
\end{suppinfo}

\bibliography{bib2}

\end{document}